\documentclass[12pt]{iopart}

\expandafter\let\csname equation*\endcsname\relax
\expandafter\let\csname endequation*\endcsname\relax
\usepackage{graphics} 
\usepackage{graphicx}
\usepackage{dcolumn}
\usepackage{bm}
\usepackage{subfigure}
\usepackage{multirow}
\usepackage{float}
\usepackage{amsmath}
\usepackage{esvect}
\usepackage{autobreak}
\usepackage{multirow,makecell}

\begin{document}
\title{Pseudorapidity distributions of charged particles in pp($\rm\overline{p}$), p(d)A and AA collisions using Tsallis thermodynamics}

\author{J.Q. Tao, M. Wang, H. Zheng*, W.C. Zhang}
\address{School of Physics and Information Technology, Shaanxi Normal University, Xi’an 710119, China}
\eads{zhengh@snnu.edu.cn}
\author{L.L. Zhu}
\address{Department of Physics, Sichuan University, Chengdu 610064, China}
\author{A. Bonasera}
\address{Cyclotron Institute, Texas A\&M University, College Station, TX 77843, USA}
\address{Laboratori Nazionali del Sud, INFN, 95123 Catania, Italy}

\begin{abstract}
The pseudorapidity distributions of charged particles measured in p+p($\rm \overline{p}$) collisions for energies ranging from $\sqrt{s_{NN}}=23.6$ GeV to 13 TeV, d+Au collisions at $\sqrt{s_{NN}}=200$ GeV, p+Pb collisions at $\sqrt{s_{NN}}= 5.02$ TeV and A+A collisions at RHIC and LHC are investigated in the fireball model with Tsallis thermodynamics. We assume that the rapidity axis is populated with fireballs following q-Gaussian distribution and the charged particles follow the Tsallis distribution in the fireball. The theoretical results are in good agreement with the experimental data for all the collision systems and centralities investigated. The collision energy and centrality dependence of the central position $y_0$ and its width $\sigma$ of the fireball distribution are also investigated. A possible application of the model to predict the charged particle pseudorapidity distributions for the system size scan program proposed recently for the STAR experiment at RHIC is proposed.
\end{abstract}
\maketitle

\section{Introduction}\label{sec1}

High energy heavy-ion collisions performed in the laboratory have been used to study the systems with hadronic or partonic degrees of freedom under extremely high temperature and density. It is commonly agreed that a new dense and hot phase of nuclear matter denoted as quark-gluon plasma (QGP), predicted by Quantum-Chromodynamics (QCD), is created where the partonic degrees of freedom come into play during the dynamical evolution of the system \cite{Wong1994, Csernai1994}. But in experiment, only the final states of produced particles can be measured, including the charged particle pseudorapidity distributions ($\frac{dN_{ch}}{d\eta}$), particle transverse momentum spectra and so on. Given those information, we are able to study the properties of QGP, particle production mechanism, particle correlations etc.. The study of the charged particle pseudorapidity distribution over a wide range of pseudorapidity ($\eta$) and its dependence on the colliding system, energy as well as centrality, are benchmark tools to constrain the models \cite{Abbas2013, Adam20162, Adam20172, Acharya2019}. It is important to  improve our understanding of the particle production mechanisms and provide insight into the partonic structure of the interacting nuclei.  At the same time, the total charged particle multiplicities and the evolution of their distributions with centrality add significant information. 

In experiments, abundant data of the pseudorapidity distributions of charged particles have been measured by different experimental collaborations \cite{Abbas2013,Adam20162,Adam20172, Acharya2019,Alner1986,Alver2011,Aamodt2010,Adam2017,Adam2016,Abe1990,Arnison1983,thome1977,Alver2009,Back2003,Back2005,Aad2014,Back20052}, ranging from the small collision systems such as p+p to large ones such as Pb+Pb at different collision energies and centralities. In order to describe $\frac{dN_{ch}}{d\eta}$, experimentalists have proposed several parameterizations and utilized them to extract the total charged particle multiplicities ($N_{ch}^{tot}$) produced in the reactions \cite{Abbas2013, Adam20162, Back2003, basu2020}. Some well established  models, such as HIJING \cite{Wang1991}, AMPT \cite{Lin2005} (with and without string melting), EPOS-LHC \cite{Pierog2015}, UrQMD \cite{Mitrovski2009}, CGC based model \cite{Albacete2011}, introduced in the field of high energy heavy-ion collisions, are also adopted to understand the pseudorapidity distributions of charged particles as well as other physics. 

In Refs. \cite{Sun2013, Li2014}, a multi-source thermal model, the four sources being projectile and target cylinders and the projectile and target leading particles, has been applied to describe the centrality dependence of the charged particle pseudorapidity distribution in Pb+Pb collisions. In Ref. \cite{Gao2015}, a new revised Landau hydrodynamic model following the same philosophy as in the multi-source thermal model is proposed to systematically study $\frac{dN_{ch}}{d\eta}$ produced in heavy-ion collisions. The system is assumed to be consisted of three sources, i.e., the central, target and projectile sources. The central source is described by the Landau hydrodynamic model and the target and projectile sources are assumed to emit particles in their rest frame isotropically. In Refs. \cite{jiang22018,jiang2016}, a 1+1 dimensional hydrodynamics model, which is analytically solvable,  is adopted by taking into account the collective motion to study the pseudorapidity distributions in both AA and pp collisions at currently available energies. In this model, the final particles have been categorized as the particles governed by the hydrodynamics and the leading particles. In Ref. \cite{jiang20182}, a relativistic viscous hydrodynamics model considering the longitudinal acceleration effect has been used to study the pseudorapidity distribution of charged particles and estimate the energy density from the most central AA collisions at RHIC and LHC. These models have different scenarios, but their results are in good agreement with the available experimental data.

Recently, the fireball model based on Tsallis thermodynamics had been successfully applied to describe the charged particle (pseudo)rapidity distributions produced in heavy-ion collisions \cite{Marques2015,Gao2017}. In this paper, we will systematically analyze the pseudorapidity distributions of charged particles produced in p+p($\rm \overline{p}$) collisions at energies ranging from $\sqrt{s_{NN}}=23.6$ GeV to 13 TeV, d+Au collisions at $\sqrt{s_{NN}}=$200 GeV, p+Pb collisions at $\sqrt{s_{NN}}=$5.02 TeV and A+A collisions including Cu+Cu, Au+Au, Pb+Pb and Xe+Xe at RHIC and LHC energies and different centralities. We will also extract the total charged particle multiplicities from the model and compare with the experimental data and study their dependence on centrality as well as collision energy.

The paper is organized as follows: In section II, we briefly introduce the fireball model with Tsallis thermodynamics for the pseudorapidity distribution. Then we extend the model to asymmetric collision system. In section III, we show the model results along with the experimental data of the pseudorapidity distributions of charged particles produced in collision systems selected. The corresponding total charged particle multiplicities are also extracted and studied. Furthermore, the collision energy and centrality dependence of the central position $y_0$ and its width $\sigma$ of the fireball distribution are also discussed. In section IV, a brief conclusion is drawn.

\begin{figure}[ht]
	\centering
	\begin{tabular}{c}
		\includegraphics[scale=0.72]{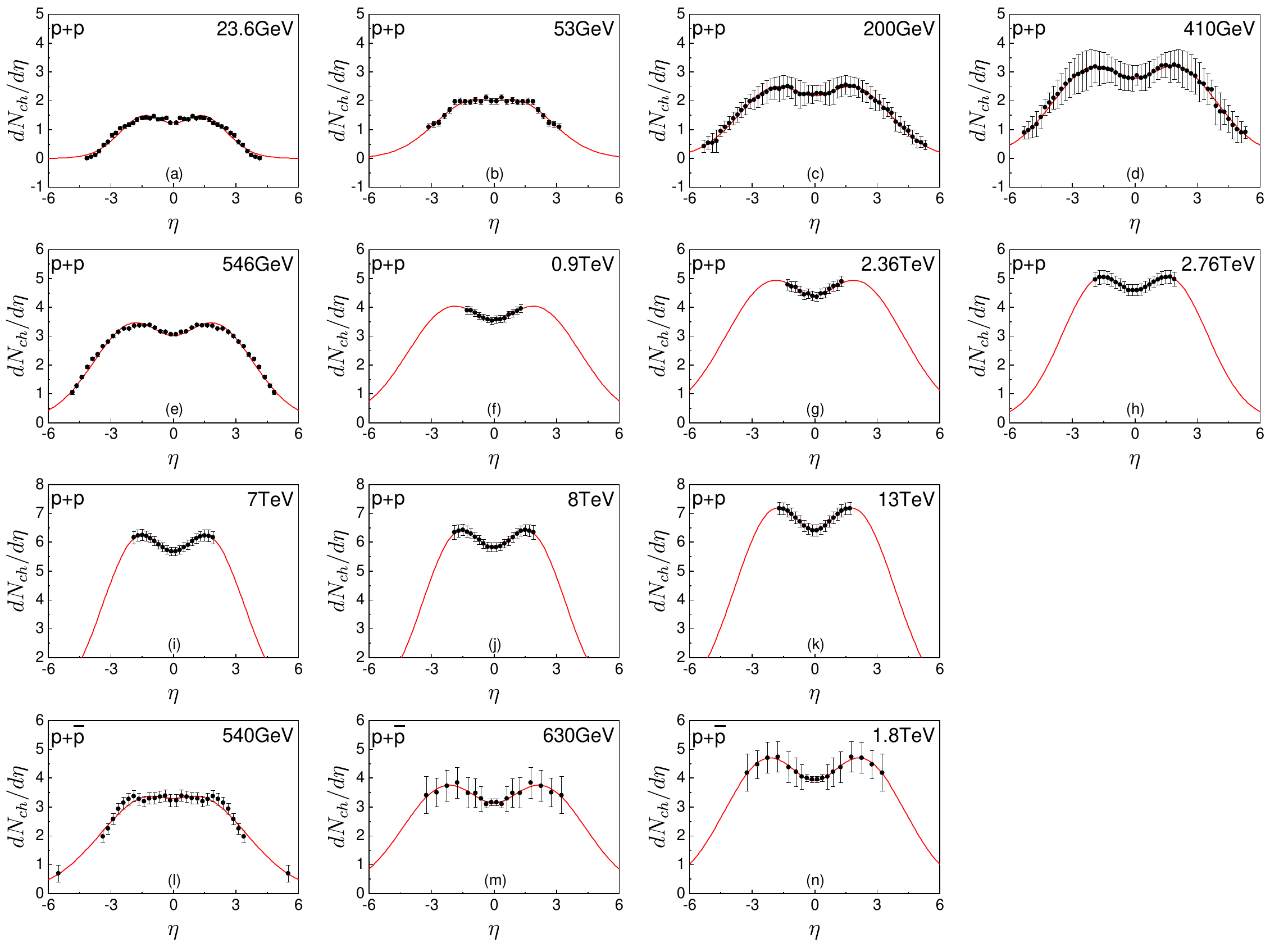}
	\end{tabular}
	\caption{(Color online) The pseudorapidity distribution of charged particles produced in proton-(anti)proton collisions for energies ranging from $\sqrt{s_{NN}}=23.6$ GeV to 13 TeV. The symbols are experimental data taken from Refs.  \cite{Alner1986,Alver2011,Aamodt2010,Adam2017,Adam2016,Abe1990,Arnison1983,thome1977}. The curves are the results from Eqs. (\ref{8}) and (\ref{10}).}
	\label{p+(p_)p}
\end{figure}

\section{Theoretical framework and formulas}\label{sec2}

Recently, the Tsallis distribution has been extensively and successfully applied to describe the particle spectra produced in pp, pA and AA collisions. In Refs. \cite{Zheng2016, Zheng2015}, it has been shown that different versions of the Tsallis distribution can reproduce the particle spectra resulting in different fitting parameters. In the framework of the fireball model with Tsallis thermodynamics \cite{Marques2015, Cleymans2012}, the transverse momentum distribution of particles is taken as
\begin{eqnarray}
		\frac{d^{2} N}{2 \pi p_{T} d p_{T} d y}=g V \frac{m_{T} \cosh y}{(2 \pi)^{3}}\left[1+(q-1) \frac{m_{T} \cosh y-\mu}{T}\right]^{-\frac{q}{q-1}},\label{1}
\end{eqnarray}
where $V$ is the volume, $g$ is the particle state degeneracy,  $m_{T}=\sqrt{m^{2}_{0}+p_{T}^{2}}$ is the transverse mass and $m_{0}$ is the particle rest mass, $y$ is the rapidity, $\mu$ is the chemical potential, $T$ is the temperature and $q$ is the entropic factor which measures the nonadditivity of the entropy. When $q=1$, the Tsallis distribution becomes the Boltzmann distribution. There are four parameters $V$, $\mu$, $T$ and $q$ in Eq. (\ref{1}) to describe the particle spectrum. To reduce the number of free parameters and simplify the problem, $\mu$ is assumed to be 0 for all the collision systems in this study. In the mid-rapidity $y\approx0$ region, Eq. (\ref{1}) can be written as
\begin{eqnarray}
	\frac{d^{2} N}{2 \pi p_{T} d p_{T} d y}=g V \frac{m_{T}}{(2 \pi)^{3}}\left[1+(q-1) \frac{m_{T}}{T}\right]^{-\frac{q}{q-1}}, \label{2}
\end{eqnarray}
which will be used to extract the parameters $q$ and $T$ by fitting the particle transverse momentum spectrum. 

In the framework of the fireball model with Tsallis thermodynamics \cite{Marques2015, Gao2017}, the rapidity axis is populated with fireballs which follow the distribution function $\nu(y_{f})$, where $y_f$ is the rapidity of the fireball. The particles emitted from the fireball follow the Tsallis distribution Eq. (\ref{1}). Therefore, the particle distribution measured can be written as
\begin{equation}
\frac{d^{2} N}{p_{T} d p_{T} d y}= \frac{N}{A} \int_{-\infty}^{\infty} \nu\left(y_{f}\right) \frac{m_{T} \cosh \left(y-y_{f}\right)}{(2 \pi)^{2}}\left[1+(q-1) \frac{m_{T} \cosh \left(y-y_{f}\right)}{T}\right]^{-\frac{q}{q-1}} d y_{f},
 \label{3}
\end{equation}
where $N$ is the total particle multiplicity and $A$ is the normalization constant that ensures
\begin{eqnarray}\label{4}
	\int_{-\infty}^{\infty} \int_{0}^{\infty} \frac{d^{2} N}{p_Td p_{T} d y} p_Td p_{T} d y=N.
\end{eqnarray}
To deduce the formula of the rapidity density distribution $\frac{dN}{dy}$ of charged particles, we integrate Eq. (\ref{3}) over the transverse momentum $p_{T}$ and obtain \cite{Marques2015, Gao2017}
\begin{eqnarray}
	\frac{dN}{dy}&=&\frac{N}{A} \int_{-\infty}^{\infty} dy_f \nu(y_f) T \left[1+m_0 (q-1)\frac{\cosh(y-y_f)}{T}\right]^{-\frac{1}{q-1}} \nonumber\\
	&&\times \frac{-(q-2)m_0^2+2m_0T \textrm{sech}(y-y_f)+2T^2 \textrm{sech}^2(y-y_f)}{4\pi^2(q-2)(2q-3)}.\label{5}
\end{eqnarray}

In order to compare with the experimental data, one should convert Eq. (\ref{5}) to pseudorapidity density distribution $\frac{dN}{d\eta}$ of charged particles because data are usually measured in pseudorapidity space. The relation between the pseudorapidity and rapidity is \cite{Wong1994}
\begin{eqnarray}
	\frac{d y}{d \eta}\left(\eta, p_{T}\right)=\sqrt{1-\frac{m_{0}^{2}}{m_{T}^{2} \cosh ^{2} y}}.\label{6}
\end{eqnarray}
Substituting Eq. (\ref{6}) into Eq. (\ref{3}), we obtain:
\begin{equation}
\frac{d^2N}{p_Tdp_Td\eta } =  \frac{N}{A}\sqrt{1-\frac{m_0^2}{m_T^2 \cosh^2 y}}\int_{-\infty}^{\infty} \nu(y_f)\frac{m_T\cosh(y-y_f)}{(2\pi)^2}\left[1+(q-1)\frac{m_T\cosh(y-y_f)}{T}\right]^{-\frac{q}{q-1}}dy_f.
	\label{7}
\end{equation}

Because of the extra term $\sqrt{1-\frac{m_0^2}{m_T^2 \cosh^2 y}}$, Eq. (\ref{7}) cannot be analytically integrated over $p_T$. Therefore, we perform the numerical integration for the  pseudorapidity distribution 
\begin{eqnarray}
	\frac{dN}{d\eta}&=& \frac{N}{A}\int_{-\infty}^{\infty}dy_f \int_0^\infty dp_T p_T   \sqrt{1-\frac{m_0^2}{m_T^2 \cosh^2 y}}  \nonumber\\
	&&\times \nu(y_f)\frac{m_T\cosh(y-y_f)}{(2\pi)^2}\left[1+(q-1)\frac{m_T\cosh(y-y_f)}{T}\right]^{-\frac{q}{q-1}}, 
	\label{8}
\end{eqnarray}

where
\begin{eqnarray}
	y=\frac{1}{2} \ln \left[\frac{\sqrt{p_{T}^{2} \cosh ^{2} \eta+m_{0}^{2}}+p_{T} \sinh \eta}{\sqrt{p_{T}^{2} \cosh ^{2} \eta+m_{0}^{2}}-p_{T} \sinh \eta}\right].
	\label{9}
\end{eqnarray}

The distribution $\nu(y_f)$ of the  fireballs has been assumed to be the summation of two q-Gaussian functions for the symmetric collision systems in Refs.\cite{Marques2015, Gao2017} as

\begin{equation}\label{10}
	\nu\left(y_{f}\right)=\frac{1}{\sqrt{2 \pi} \sigma}\left[1+\left(q^{\prime}-1\right) \frac{\left(y_{f}-y_{0}\right)^{2}}{2 \sigma^{2}}\right]^{-\frac{1}{q^{\prime}-1}}+\frac{1}{\sqrt{2 \pi} \sigma}\left[1+\left(q^{\prime}-1\right) \frac{\left(y_{f}+y_{0}\right)^{2}}{2 \sigma^{2}}\right]^{-\frac{1}{q^{\prime}-1}},
\end{equation}
where $y_0$ and $\sigma$ are fitting parameters referring to the central position and its width of the fireball distribution respectively, which are determined by the experimental data.  $q^{\prime}$ is another fitting parameter which can be different from $q$ for the particle transverse momentum spectrum Eq. (\ref{1}) while $q'=q$ is assumed in Refs.  \cite{Marques2015,Gao2017}. Since the symmetric and asymmetric collision systems will be considered in this work, we extend the fireball model with Tsallis thermodynamics to the asymmetric collision system and modify $\nu(y_f)$ as
\begin{equation}\label{11}
\nu\left(y_{f}\right)=\frac{1}{\sqrt{2 \pi} \sigma_{01}}\left[1+\left(q^{\prime}-1\right) \frac{\left(y_{f}-y_{01}\right)^{2}}{2 \sigma_{01}^{2}}\right]^{-\frac{1}{q^{\prime}-1}}+\frac{x}{\sqrt{2 \pi} \sigma_{02}}\left[1+\left(q^{\prime}-1\right) \frac{\left(y_{f}+y_{02}\right)^{2}}{2 \sigma_{02}^{2}}\right]^{-\frac{1}{q^{\prime}-1}}, 
\end{equation}
where $y_{0i}$ and $\sigma_{0i}$ ($i=1, 2$) are the central positions and their widths of the fireball distributions. $x$ is the relative weight introduced for the asymmetric collision system. For the symmetric collision system, $y_{01}=y_{02}$, $\sigma_{01}=\sigma_{02}$ and $x=1$ and Eq. (\ref{10}) is recovered. In analogy to Refs. \cite{Marques2015,Gao2017}, $q'=q$ is assumed.
\begin{figure}[ht]
	\centering
	\includegraphics[scale=0.5]{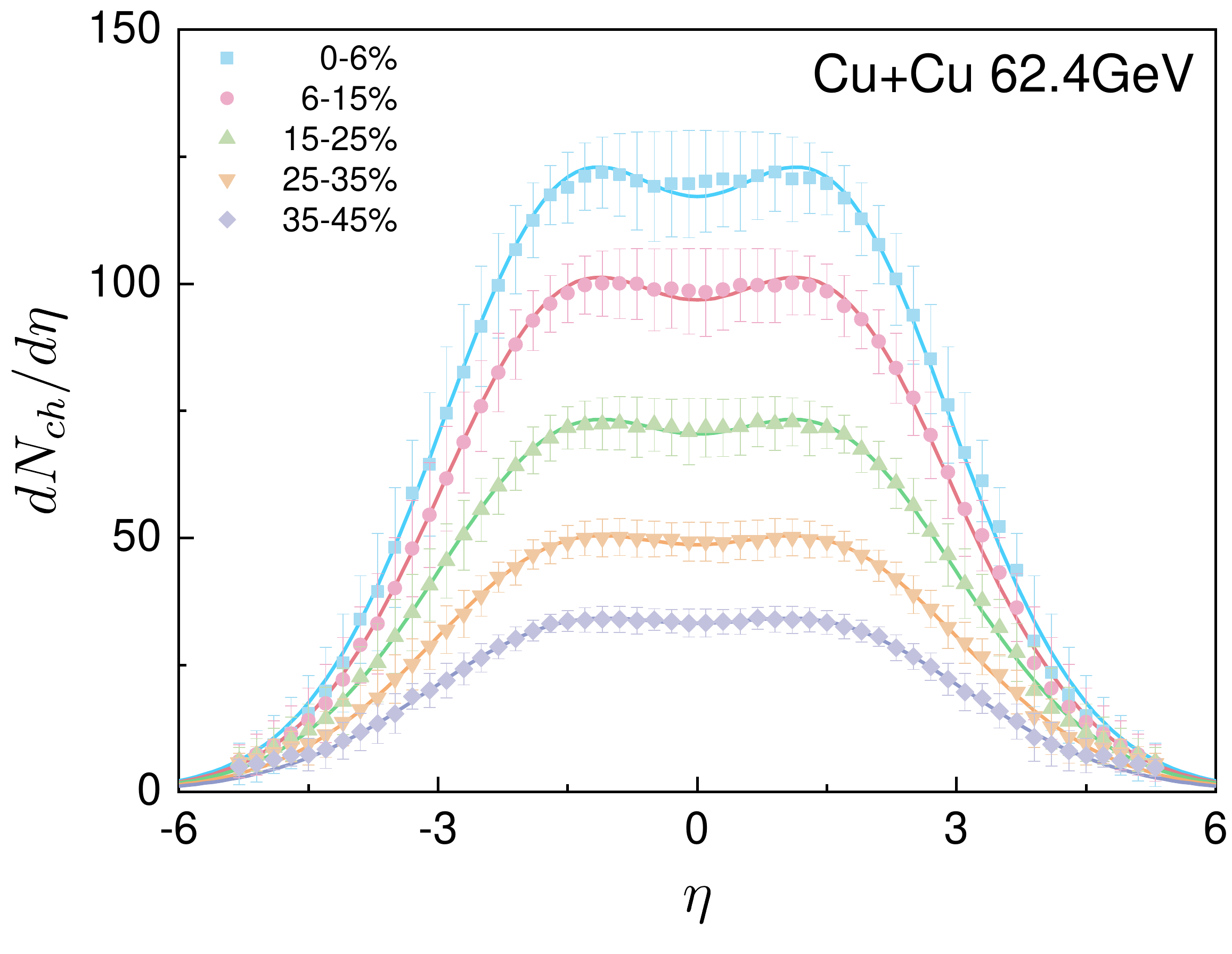}
	\caption{(Color online) The pseudorapidity distribution of charged particles produced in Cu+Cu collisions at  $\sqrt{s_{NN}}=62.4$ GeV for different centralities. The symbols are experimental data taken from Ref. \cite{Alver2009}. The curves are the results from Eqs. (\ref{8}) and (\ref{10}).}\label{Cu+Cu}
\end{figure}

\begin{figure}[ht]
	\centering
	\includegraphics[scale=0.5]{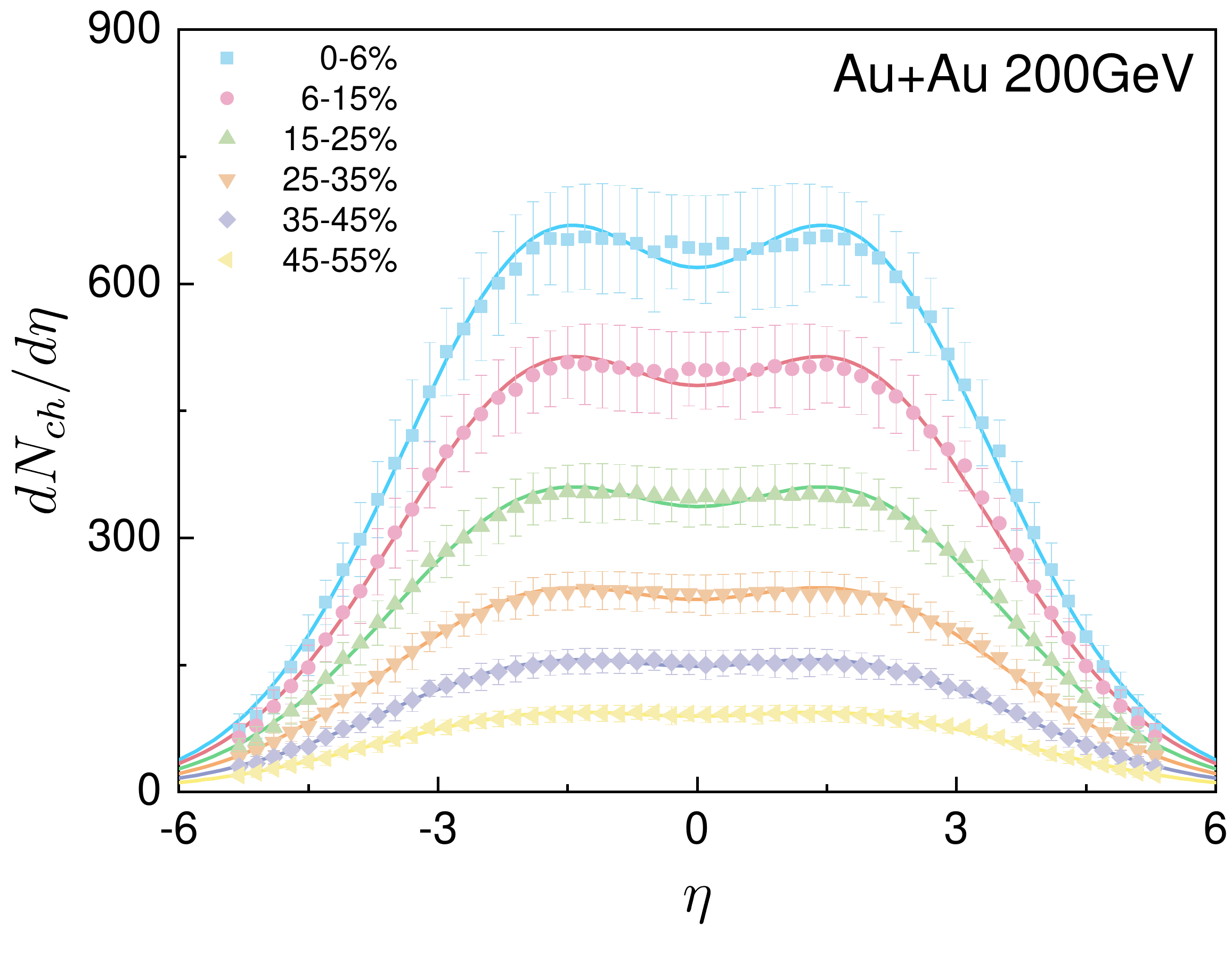}
	\caption{(Color online) The pseudorapidity distribution of charged particles produced in Au+Au collisions at  $\sqrt{s_{NN}}=200$ GeV for different centralities. The symbols are experimental data taken from Ref. \cite{Back2003}. The curves are the results from Eqs. (\ref{8}) and (\ref{10}).}
	\label{Au+Au}
\end{figure}

\begin{figure}[ht]
	\centering
	\includegraphics[scale=0.5]{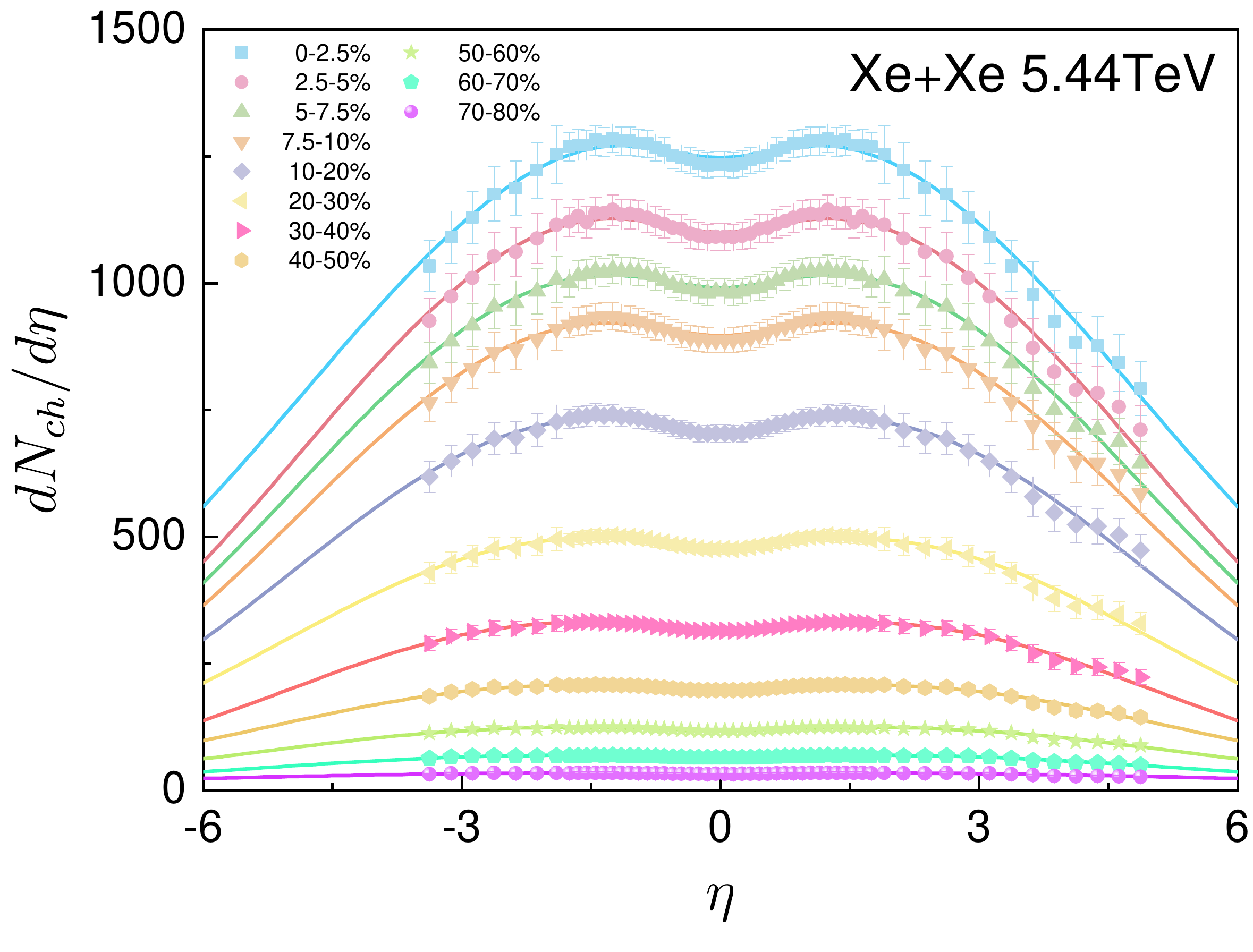}
	\caption{(Color online) The pseudorapidity distribution of charged particles produced in Xe+Xe collisions at  $\sqrt{s_{NN}}=5.44$ TeV for different centralities. The symbols are experimental data taken from Ref. \cite{Acharya2019}. The curves are the results from Eqs. (\ref{8}) and (\ref{10}).}
	\label{Xe+Xe}
\end{figure}

\begin{figure}[ht]
	\centering
	\includegraphics[scale=0.5]{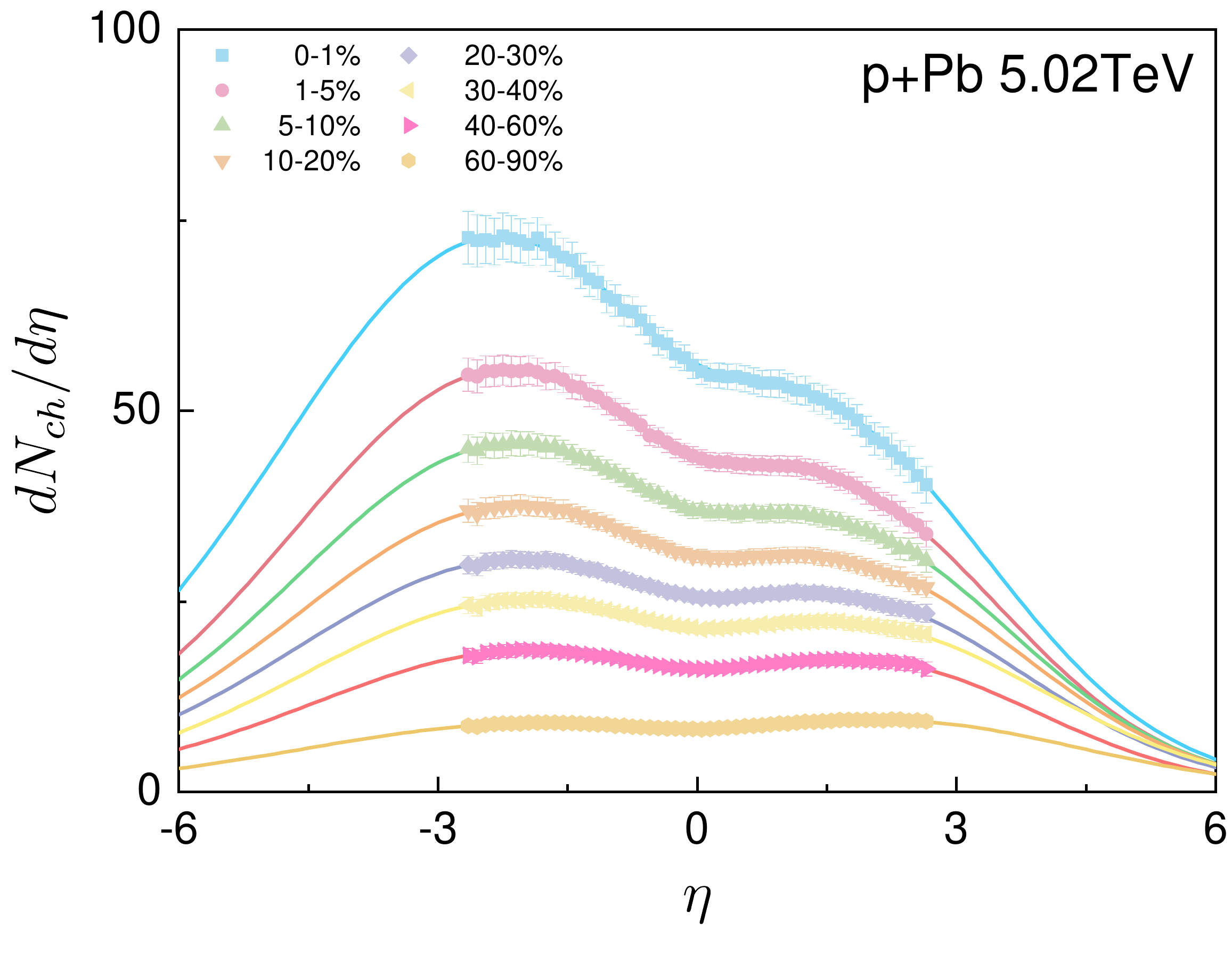}
	\caption{(Color online) The pseudorapidity distribution of charged particle produced in p+Pb collisions at  $\sqrt{s_{NN}}=5.02$ TeV for different centralities. The symbols are experimental data taken from Ref. \cite{Aad2014}. The curves are the results from Eqs. (\ref{8}) and (\ref{11}).}
	\label{p+Pb}
\end{figure}

\section{Results}
As discussed in Sec. \ref{sec2}, we need to extract the parameters $q$ and $T$ by fitting the particle transverse momentum spectra at the mid-rapidity, given by the experimental collaborations \cite{Adam2016,ptAdam2003,ptKhachatryan2010,ptAcharya2018,ptAdam2015,ptAad2016,ptArnison1982,ptAbe1988,ptAdamczyk2017,ptBack2005,ptAdler2002,ptBack2004,ptAlver2006,ptBack2003,ptAdam2016,ptAcharya2019}, using Eq. (\ref{2}).  We had pointed out that the single Tsallis distribution only can reproduce the particle spectra at low and intermediate $p_T$ region, where most of the particles are produced, in Pb+Pb and Xe+Xe collisions at LHC. Therefore, a $p_T$ cut at 7.5 GeV/c is applied for both collision systems when we fit the charged particle spectra. Since there are plenty of works to study the particle spectra with Tsallis distribution \cite{Zheng2016, Zheng2015, Zhengprd2015, pt1,pt2,pt3,pt4,pt5}, we will not show the figures here. Then we can use these two parameters and the fireball model with Tsallis thermodynamics, Eqs. (\ref{8}) and (\ref{11}) (or (\ref{10}) for the symmetric collision systems) to study the pseudorapidity distribution of charged particles for the system investigated.

In Fig.\ref{p+(p_)p}, the results of the pseudorapidity distribution of charged particles produced in proton-(anti)proton collisions, using the fireball model, from different experiments with collision energies ranging from 23.6 GeV up to 13 TeV, as well as the corresponding experimental data are shown. As one can see, the experimental results are well reproduced, analogous to the results shown in Ref. \cite{Marques2015} but with more collision energies. One should notice that the scale of the charged particle pseudorapidity density distribution $\frac{dN_{ch}}{d\eta}$ is different for each row in Fig. \ref{p+(p_)p} in order to better visualize the results. One can observe that a valley at $\eta=0$ is developed and its depth increases with the collision energy for both p+p and p+$\rm\overline{p}$ collisions. This valley is commonly related to partial transparency and limited stopping power of the colliding (anti)proton \cite{Wolschin2015}. A difference of the pseudorapidity distributions between p+p collisions and p+$\rm\overline{p}$ collisions that the valley of the pseudorapidity distribution is developed at higher collision energy in p+$\rm\overline{p}$ collisions can be easily noticed. This may reflect the fact that p and $\rm\overline{p}$ can annihilate each other when they collide while p and p cannot, so this should take off more p$\rm\overline{p}$ from zero pseudorapidity. One also can notice that the pseudorapidity distributions only cover a very limited $\eta$ range at some collision energies, e.g., 0.9 TeV of p+p collisions, because of the limited acceptance of the detectors in the experiment. This results in the large uncertainties of the central position values $y_0$ and its widths $\sigma$ of the fireball distributions. Therefore, we will not adopt these collision energies to explore the energy dependence of the central position and its width of the fireball distribution discussed later in this paper.

In Figs. \ref{Cu+Cu}, \ref{Au+Au}, \ref{Xe+Xe}, the results of the pseudorapidity distributions of charged particles produced in AA collisions, i.e., Cu+Cu collisions at $\sqrt{s_{NN}}=62.4$ GeV, Au+Au collisions at $\sqrt{s_{NN}}=200$ GeV, Xe+Xe collisions at $\sqrt{s_{NN}}=5.44$ TeV, at different centrality bins are shown.  As one can see, the experimental results at different centralities are well reproduced. We also studied the other nucleus-nucleus collision systems, i.e., Cu+Cu collisions at $\sqrt{s_{NN}}=200$ GeV, Au+Au collisions at $\sqrt{s_{NN}}=19.6, 62.4, 130$ GeV and Pb+Pb collisions at $\sqrt{s_{NN}}=2.76, 5.02$ TeV,  at RHIC and LHC. Similar results are found. Actually, the results for Au+Au collisions at $\sqrt{s_{NN}}=130, 200$ GeV, Pb+Pb collisions at $\sqrt{s_{NN}}=2.76$ TeV and the prediction for the central collisions of Pb+Pb at $\sqrt{s_{NN}}=5.02$ TeV were shown in Ref. \cite{Gao2017}. It is worth noting that the peak and valley structure of the charged particle pseudorapidity distributions produced in AA collisions are typically shallower and wider than those produced in proton-(anti)proton collisions at the similar collision energies. This can be associated with the dense medium produced in AA collisions and its evolution dynamics.

In Fig.\ref{p+Pb}, we studied the asymmetric collision system p+Pb at $\sqrt{s_{NN}}=5.02$ TeV. Different from the symmetric collision case, the pseudorapidity distribution is no longer symmetric respect to $\eta=0$ but obviously has forward/backward asymmetry that more charged particles are produced in the p side than the Pb side. The extended fireball model, Eqs. (\ref{8}) and (\ref{11}), is adopted to take into account asymmetric fact in this case. The results of the pseudorapidity distributions of charged particles for eight centrality bins are shown. A good agreement between the model results and the experimental data is shown. One can see that the pseudorapidity distribution of the charged particles becomes more symmetric from central to peripheral collisions because the peripheral collisions are more like the p+p collisions for the asymmetric collision system. Similar behavior is also observed in d+Au collisions at $\sqrt{s_{NN}}=200$ GeV \cite{Back20052}.

\begin{figure}[ht]
	\centering
	\includegraphics[scale=0.5]{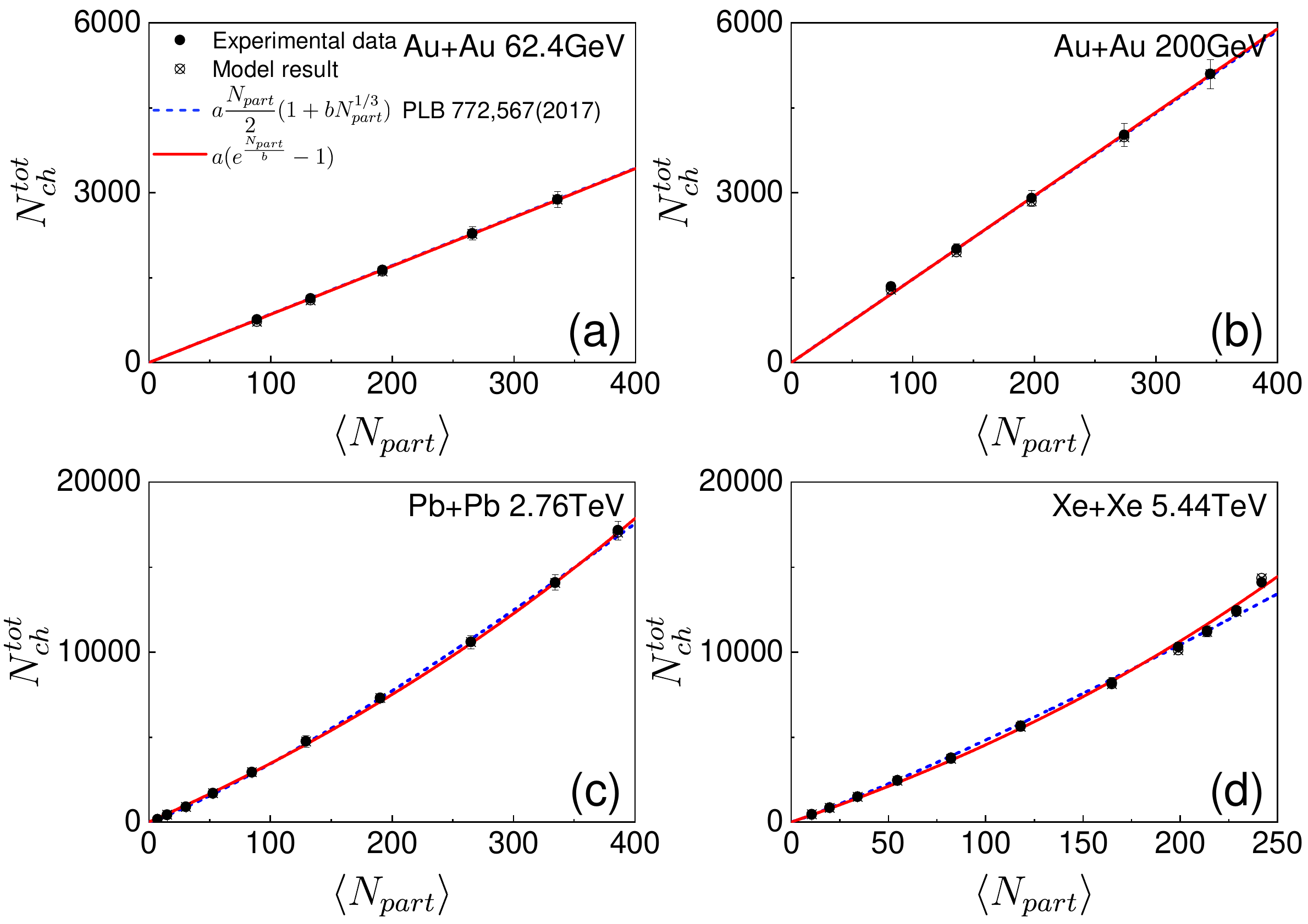}
	\caption{(Color online) The total charged particle multiplicity produced in Au+Au collisions at $\sqrt{s_{NN}}=62.4, 200$ GeV, Pb+Pb at $\sqrt{s_{NN}}=2.76$ TeV and Xe+Xe at  $\sqrt{s_{NN}}=5.44$ TeV versus the number of participants. Experimental data (solid circles) are taken from Refs. \cite{Abbas2013,Adam20162,Acharya2019, Alver2011}. The open circles with a cross inside are the model results. The lines are the fitting results  and their functions are indicated in the legend.}
	\label{totalN}
\end{figure}
\begin{figure}[ht]
	\centering
	\includegraphics[scale=0.85]{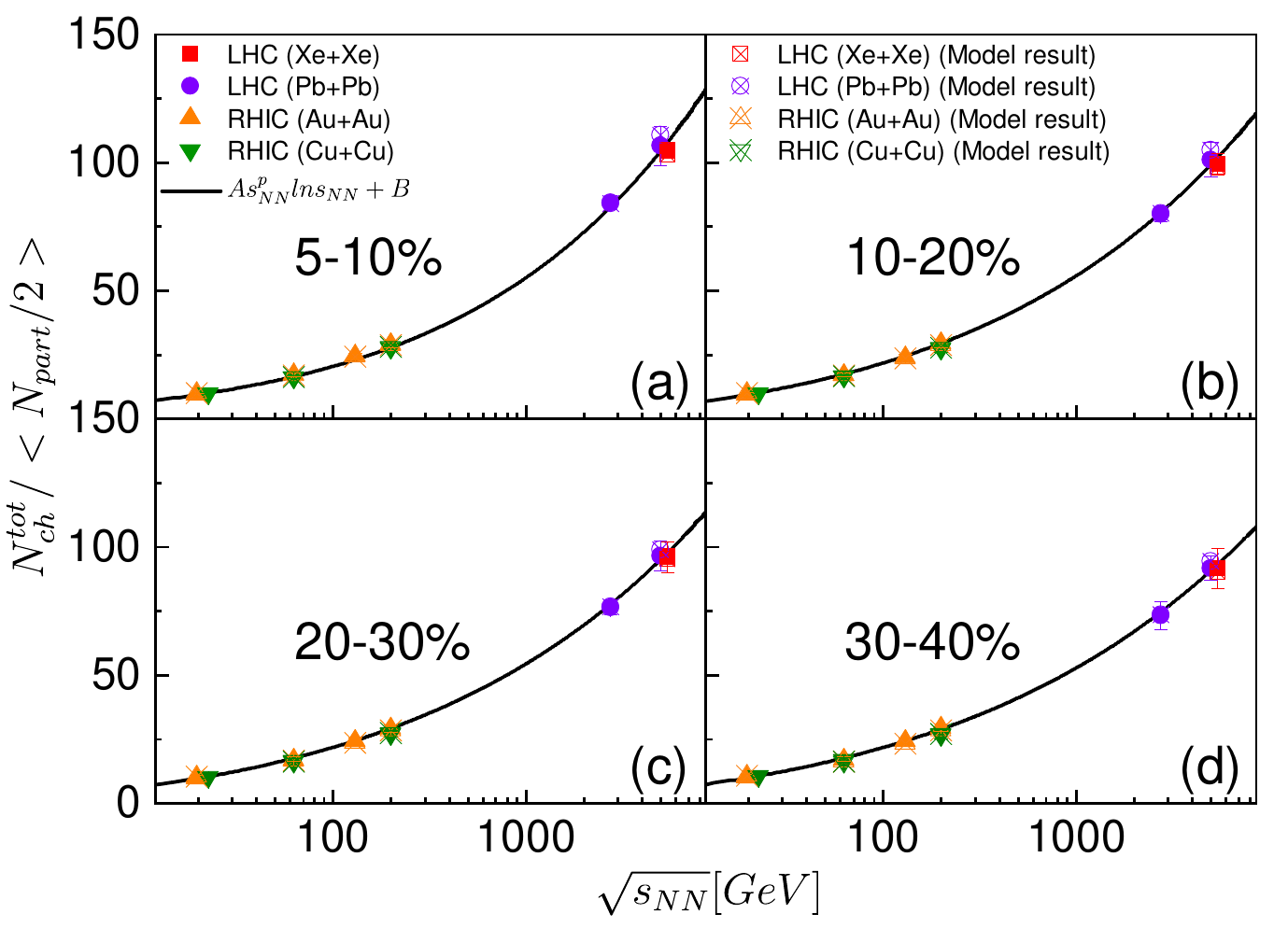}
	\caption{(Color online) The total charged particle multiplicity per participant pair produced in AA collisions at different centralities versus $\sqrt{s_{NN}}$. The experimental data are from RHIC (Cu+Cu and Au+Au) \cite{Alver2011} and LHC (Pb+Pb and Xe+Xe) \cite{Abbas2013,Adam20162,Adam20172,Acharya2019}. The curves are the fitting results and the corresponding function is adopted from Ref. \cite{Abbas2013}.}
	\label{totalNNpart}
\end{figure}

\begin{figure}[ht]
	\centering
	\includegraphics[scale=0.5]{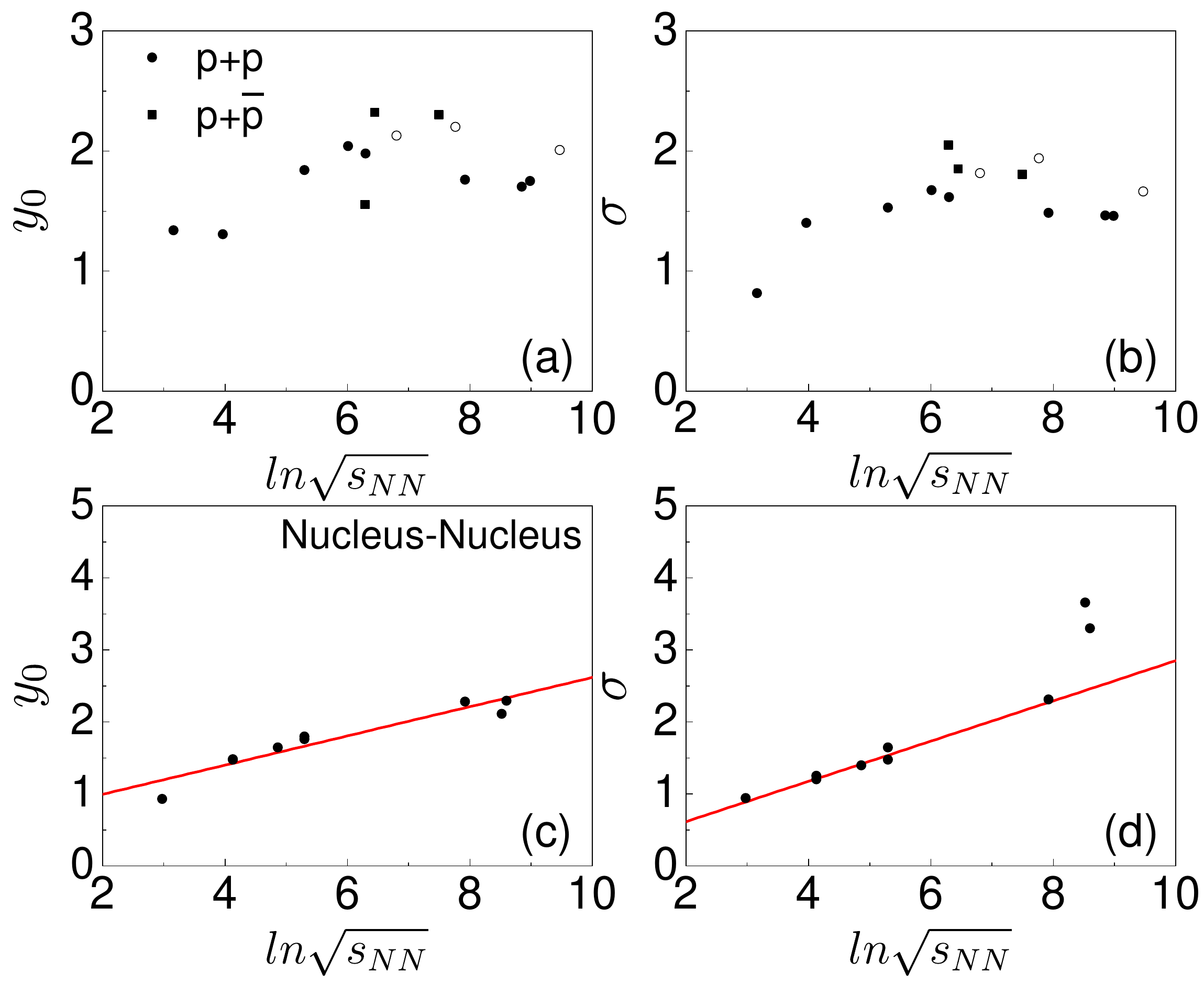}
	\caption{(Color online) The collision energy dependence of the central position $y_0$ of the fireball distribution and its width $\sigma$ in proton-(anti)proton collisions (a) and (b), and the most central nucleus-nucleus collisions (c) and (d). The open circles in (a) and (b) are the model results for the p+p collisions without two humps structure in their pseudorapidity distributions because of the detector acceptance. The AA collision systems  in (c) and (d) are Au+Au(0-6\% 19.6 GeV), Au+Au(0-6\% 62.4 GeV), Au+Au(0-6\% 130 GeV), Au+Au(0-6\% 200 GeV), Cu+Cu(0-6\% 62.4 GeV), Cu+Cu(0-6\% 200 GeV), Pb+Pb(0-5\% 2.76 TeV), Pb+Pb(0-5\% 5.02 TeV), Xe+Xe(0-5\% 5.44 TeV). The solid lines are the linear fitting results. }
	\label{parameter(sqrt(s))}
\end{figure}

\begin{figure}[ht]
	\centering
	\includegraphics[scale=1.7]{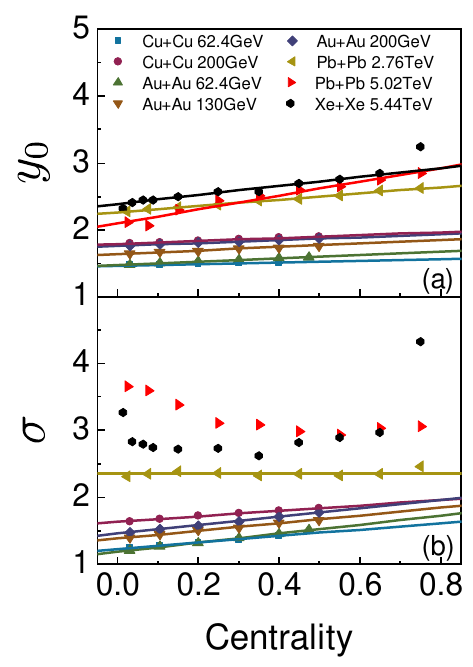}
	\caption{(Color online) The centrality dependence of the $y_0$ and $\sigma$ in Cu+Cu collisions at $\sqrt{s_{NN}}=62.4, 200$ GeV, Au+Au collisions at $\sqrt{s_{NN}}=62.4, 130, 200$ GeV, Pb+Pb collisions at $\sqrt{s_{NN}}=2.76, 5.02$ TeV and Xe+Xe collisions at $\sqrt{s_{NN}}=5.44$ TeV. The solid lines are the linear fitting results.}
	\label{parameter(centrality)}
\end{figure}

Now we turn to analyze the relation between the centrality (collision energy) and the total charged particle multiplicity produced in different collision systems. By integrating Eq. (\ref{8}) over $\eta$, the total number of charged particles produced can be extracted from the fireball model. To compare with the experimental data, the $\eta$ cut has been adopted same as the experiments when available.  Here we only show the results for AA collisions. For the pp($\rm\overline{p}$) collisions, we refer the results to Ref. \cite{Acharya2019}. While for the p(d)A collisions, we do not have enough data to conduct the analysis. In Fig. \ref{totalN}, we plot the total charged particle multiplicity versus the number of participants for the selected collision systems. As one can see that the model results are almost indistinguishable from the experimental data. Similar results are also obtained for the other collision systems investigated in this paper. The analytical function (see the legend) adopted from Ref. \cite{Adam20172} has been used to fit the total charged particle multiplicity as a function of $N_{part}$, shown by the dashed lines in the figure. This function can well reproduce all the experimental data except Xe+Xe collisions where there is a deviation for the data at central collisions. We proposed an analytical function $a(e^{N_{part}/b}-1)$, where $a$ and $b$ are fitting parameters, and it works for all the collision systems. According to the analytical functions, we cannot distinguish whether the total charged particle multiplicity increases with the number of participants exponentially or with a power-law. 

In Fig. \ref{totalNNpart}, we investigate the total charged particle multiplicity per participant pair produced in AA collisions at different centralities as a function of the collision energy. One can see that the model results are in good agreement with the experimental data which is not surprising as one has seen it from Fig. \ref{totalN}. It is found that the analytical function adopted from Ref. \cite{Abbas2013}, applied only to the central collision data, can describe the experimental data well for different centralities. The fitting power $p$ decreases from central to peripheral collisions, which qualitatively reflects the fact that the dense medium created in the more central collisions causes more collisions and leads to more charged particle production per participant pair.

\renewcommand\arraystretch{1.5}
\begin{table}
	\centering
	\caption{The linear fit of $y_0$ and $\sigma$ versus centrality represented by c.}
	\begin{tabular}{>{\centering}p{1.5cm}>{\centering}p{1.7cm}>{\centering}p{2.4cm}p{2.4cm}<{\centering}}
		\hline
		System&$\sqrt{s_{NN}}$&$y_0$&$\sigma$\\
		\hline
		\multirowcell{2}{Cu+Cu}&62.4 GeV&$0.11c+1.47$&$0.48c+1.23$\\
		&200 GeV&$0.22c+1.79$&$0.41c+1.63$\\
		\hline
		\multirowcell{3}{Au+Au}&62.4 GeV&$0.25c+1.47$&$0.68c+1.18$\\
		&130 GeV&$0.26c+1.64$&$0.57c+1.38$\\
		&200 GeV&$0.22c+1.76$&$0.63c+1.46$\\
		\hline
		\multirowcell{2}{Pb+Pb}&2.76 TeV&$0.47c+2.26$&-\\
		&5.02 TeV&$1.03c+2.10$&-\\
		\hline
		\multirowcell{1}{Xe+Xe}&5.44 TeV&$0.93c+2.34$&-\\
		\hline
	\end{tabular}\label{c1}
\end{table}

We also analyze the collision energy (centrality) dependence of the central position $y_0$ and its width $\sigma$ of the fireball distribution. For the same reason as stated before, the analysis of the collision energy dependence on $y_0$ and $\sigma$ for pA collision systems is not conducted. In Fig. \ref{parameter(sqrt(s))}(a) and (b), the results of $y_0$ and $\sigma$ versus collision energy ($\ln \sqrt{s_{NN}}$) of proton-(anti)proton collisions are shown respectively. For pp collisions, the pseudorapidity distributions of charged particles do not show the two humps structure because of the limited detector acceptance of the experiments for some energies, i.e., 0.9 TeV, 2.36 TeV, 13 TeV, see Fig. \ref{p+(p_)p}. Therefore, we show the model results, which are not well constrained by the experimental data for these energies, with open circles to see the trend.  The solid circles show that both $y_0$ and $\sigma$ increase with collision energy to a critical value and then saturate to a certain value. Although there are only three collision energies for p$\rm\overline{p}$ collisions, the same trend is observed (solid squares).  In Fig. \ref{parameter(sqrt(s))}(c) and (d), we show the model results for the most central AA collisions. Different from pp$(\rm\overline{p})$, both $y_0$ and $\sigma$ show a linear dependence of collision energy for the energies investigated which demonstrates our conjecture made in Ref. \cite{Gao2017}. A linear fit has been conducted and shown in the figure respectively. For $\sigma$,  we only do the linear fit up to 2.76 TeV. For the other centralities (We put the closest centralities together because the centrality classification is not exactly the same between RHIC and LHC), the similar behavior has been observed.

In Fig. \ref{parameter(centrality)}, the centrality dependence of the $y_0$ and $\sigma$ are shown for AA collisions. A nice linear relation between $y_0$ and centrality is found for the collision energy up to 2.76 TeV. While for the reactions Pb+Pb at $\sqrt{s_{NN}}=5.02$ TeV and Xe+Xe at $\sqrt{s_{NN}}=5.44$ TeV, the linear behavior is not so great. We performed linear fits shown with the lines in Fig. \ref{parameter(centrality)}(a). The fitting parameters are listed in Table. \ref{c1}. The positive slope indicates that the stopping power is decreasing from central to peripheral collisions because there are more collisions and dissipation processes at central collisions comparing with peripheral collisions. The linear behavior of the width $\sigma$ versus centrality is also observed at RHIC. The fitting results are shown by the lines Fig. \ref{parameter(centrality)}(b) and fitting parameters are listed in Table. \ref{c1}. For the collisions at LHC, $\sigma$ shows different dependence on centrality. For Pb+Pb at  $\sqrt{s_{NN}}=2.76$ TeV, $\sigma$ tends to be a constant for all the centralities and a constant line is plotted. While the collision energy goes up, i.e., Pb+Pb at $\sqrt{s_{NN}}=5.02$ TeV and Xe+Xe at $\sqrt{s_{NN}}=5.44$ TeV,  $\sigma$ shows a U shape. We also notice a different energy dependence behavior of $\sigma$ in Fig. \ref{parameter(sqrt(s))}(d). We do not have an explanation for these different behaviors. From the results shown in Fig. \ref{parameter(centrality)}, one also can conclude that the collision system size dependence of the $y_0$ and $\sigma$ at the RHIC is weak by comparing the results obtained from Cu+Cu and Au+Au collisions at the same collision energy. Combining the results shown in Figs. \ref{totalNNpart} and \ref{parameter(sqrt(s))}, it becomes possible that one can use the fireball model to predict the charged particle pseudorapidity distributions for the system size scan program proposed recently for the STAR experiment at RHIC. A machine learning technique can provide powerful tool to this application of the fireball model.

\section{Conclusions}
In this paper, the pseudorapidity distributions of charged particles produced in different collision systems at  RHIC and LHC energies have been studied in the fireball model based on Tsallis thermodynamics. We have extended the model to the asymmetric collision systems. The theoretical results are in good agreement with the experimental data for all the collision systems investigated. The total charged particle multiplicities are extracted from model and compared with experimental data. Whether the total charged particle multiplicity increases with the number of participants exponentially or with a power-law can not be distinguished. We also investigate the collision energy and centrality dependence of the central position $y_0$ and its width $\sigma$ of the fireball distribution. In AA collisions, both the $y_0$ and $\sigma$ have a linear relation with the collision energy and centrality, in particular at RHIC energy, respectively. With the results found in this paper, a possible application of the fireball model with Tsallis thermodynamics to predict the charged particle pseudorapidity distributions for the system size scan program proposed recently for the STAR experiment at RHIC is suggested.

\section*{Acknowledgments}

This work is supported by the National Natural Science Foundation of China (Grants No. 11905120), the Fundamental Research Funds for the Central Universities No. GK201903022 and No. GK202003019 and Natural Science Basic Research Plan in Shaanxi Province of China (program No. 2020JM-289). 

\section*{References}
	
\end{document}